\documentclass{eptcs}
\providecommand{\event}{TFPIE 2014} 
\usepackage{breakurl}             
\usepackage[noxcolor]{pstricks}
\usepackage{pst-node}
\usepackage{graphicx}
\usepackage{subfigure}
\usepackage{stfloats}
\usepackage{multido}
\usepackage{setspace}
\usepackage{textcomp}

\definecolor{LightBlue}{rgb}{0.68,0.85,0.9}

\definecolor{PaleGreen}{rgb}{0.88,1,0.88}

\title{Functional Automata \\ \large{Formal Languages for Computer Science Students}}
\author{Marco T. Moraz\'an
\institute{Seton Hall University}
\email{morazanm@shu.edu}
\and
Rosario Antunez
\institute{City College of New York}
\email{mrantunez@gmail.com}
}
\def\titlerunning{Functional Automata}
\def\authorrunning{M. T. Moraz\'an \& R. Antunez}
\begin{document}
\maketitle

\begin{abstract}
An introductory formal languages course exposes advanced undergraduate and early graduate students to automata theory, grammars, constructive proofs, computability, and decidability. Programming students find these topics to be challenging or, in many cases, overwhelming and on the fringe of Computer Science. The existence of this perception is not completely absurd since students are asked to design and prove correct machines and grammars without being able to experiment nor get immediate feedback, which is essential in a learning context. This article puts forth the thesis that the theory of computation ought to be taught using tools for actually building computations. It describes the implementation and the classroom use of a library, \textsf{FSM}, designed to provide students with the opportunity to experiment and test their designs using state machines, grammars, and regular expressions. Students are able to perform random testing before proceeding with a formal proof of correctness. That is, students can test their designs much like they do in a programming course. In addition, the library easily allows students to implement the algorithms they develop as part of the constructive proofs they write. Providing students with this ability ought to be a new trend in the formal languages classroom.
\end{abstract}

\section{Introduction}
An introductory formal languages course exposes advanced undergraduate and early graduate students to automata theory, grammars, constructive proofs, computability, and decidability. Programming-oriented students find these topics to be challenging or, in many cases, overwhelming and on the fringe of Computer Science. The existence of this perception is not completely absurd since students are asked to design and prove correct machines/grammars without being able to experiment nor get immediate feedback, as they do in any programming course, which is essential in a learning context and needs to be timely (the sooner the better) \cite{Race}. Computer Science students are accustomed to immediate feedback as they routinely use compilers and interpreters that provide error messages and warnings. Moreover, machines/grammars are representations of programs. Thus, in essence, designing machines and grammars without being able to test them goes against the grain of what students have learned in programming courses. The same holds true when arguing that a problem is decidable. In our experience, the inability to implement state machines and grammars results in inexperienced students submitting solutions that are incorrect even when an exercise is relatively simple. Experimentation and immediate feedback are key components to learning that are seldom found in the automata theory classroom.

Rarely do formal languages textbooks (e.g., \cite{Lewis,Martin,Sipser}), unlike programming textbooks, offer any software infrastructure for students to experiment with their ideas and designs. This omission has an impact on how well students understand constructive proofs. In essence, a constructive proof spells out an algorithm to build a machine/grammar. For example, algorithms are outlined by the proof that nondeterministic finite-state machines and deterministic finite-state machines are equivalent and by the proof that context-free languages are closed under union. Computer Science students are taught that algorithms need to be implemented and tested. This is especially true when the student has an intuition of how an algorithm ought to proceed, but is uncertain of all the details--the typical case when a student starts thinking about a constructive proof. Without feedback obtained from implementation and testing, it is common for students to try to prove the correctness of a machine/grammar that is incorrect. This leads to their work being marked down by an instructor and, in turn, to frustration and apathy for the material. The reader can contrast this scenario with a programming course in which the student can experiment with an implementation. Formal languages textbooks that reference software support (e.g., \cite{Linz}) fail to integrate the use of the software as a part of the textbook. Instead, it is left as a recommendation. More worrisome, however, is that the support software is not easily extendible by students to integrate the algorithms they develop as part of their own constructive proofs. In other words, students are required to design algorithms that are not to be implemented.

Given that it is reasonable for Computer Science students to be able to experiment with and get feedback on the algorithms they develop, a decision must be made as to how to engage programming students. One solution, of course, is to have students implement the algorithms from scratch in their favorite programming language without providing any software support. Although possible, this approach is likely to be too time-consuming within the confines of one semester. A more reasonable approach is to provide students with a library (or a programming language) that allows them to quickly implement and test\footnote{Tests provide a form of immediate feedback that either confirms or challenges the understanding of a student.} the machines/grammars they design and the algorithms they develop. This article describes a \textsf{Racket} library, \textsf{FSM}, designed to provide students with the opportunity to implement and to test their designs using finite-state machines, regular expressions, regular grammars, pushdown automata, context-free grammars, Turing machines, and context-sensitive grammars much like they experiment when writing programs in a programming course. In addition, the \textsf{FSM} library easily allows students to implement the algorithms they develop as part of the constructive proofs they write. In essence, this article puts forth that the theory of computation ought to be taught using tools for actually building computations. Providing this ability should be a new trend in the formal languages classroom and in the development of software to support such courses. This software support allows students to receive immediate feedback on their work and, thus, the learning experience is enhanced. Furthermore, it is likely to provide significant help to instructors that commonly need to grade contrived machine designs for which determining correctness is difficult. This approach ties in nicely with other topics in a Computer Science curriculum, such as programming languages and compilers, by exposing students to programming with the data representation of programs.

The library provides the ability to replace pencil-and-paper designs with programming by making it an integral part of an Automata Theory course. Students use the library to implement and test all their designs and algorithms. The library has been implemented in \textsf{Racket}, because the resulting machines/grammars descriptions are concise.  Furthermore, most of our students are familiar with the language syntax and with list-based processing. The reader can note that the library is also useful for students not familiar with \textsf{Racket} as only a minimal amount of syntax needs to be learned. Knowledge of functional programming is not a prerequisite to use \textsf{FSM}.

The article is organized as follows. Section 2 highlights the interface and the implementation of the \textsf{FSM} library to provide the framework needed by any reader that desires to implement the library using a different programming language. Section 3 provides examples of how the library has been used in practice and highlights the usefulness of immediate feedback in the formal languages classroom. Section 4 describes and contrast \textsf{FSM} with related work. Finally, Section 5 presents concluding remarks and directions for future work.

\section{Interface and Implementation}
The \textsf{FSM} library presents the user with a generic interface to construct and manipulate state machines and grammars. The interface is presented to facilitate its reproduction in any modern programming language. Constructors are divided into two categories: primitive constructors and transformers. Primitive constructors build a state machine or a grammar from a formal description provided by the programmer. Transformers build a state machine or a grammar from existing machines or grammars exploiting algorithms obtained from constructive proofs. Observers are divided into three categories: accessors, applicators, and testers. Accessors return a specified component used to build a grammar or a state machine. Applicators apply a given machine or grammar to a word. Testers allow for machines and grammars to be tested with words provided by the programmer or with randomly generated words by the software. The latter two provide students with immediate feedback on the validity (not the verification) of their designs and implementations.

\subsection{State Machines}
A state machine (sm) is either:
 \begin{enumerate}
   \item Deterministic finite-state machine (\textsf{dfa})
   \item Nondeterministic finite-state machine (\textsf{ndfa})
   \item Pushdown automata (\textsf{pda})
   \item Turing machine (\textsf{tm})
 \end{enumerate}

Every state machine has the user explicitly provide a finite set of states ($S$), a tape alphabet ($\Sigma$), a starting state ($s$), a set of final states ($F$), and a set of transition rules ($\delta$). The transition rules must describe a function for a \textsf{dfa} while simply a relation for the other machines. Finally, a \textsf{pda} also requires a stack alphabet ($\Gamma$). $|M|$ denotes the representation of a regular expression or a machine/grammar of type $M$. The corresponding primitive constructors for machines with explicit rules have the following signatures\footnote{$\epsilon$ represents the empty string and $\sqcup$ represents a blank tape space.}:
\begin{enumerate}
   \item \textbf{make-dfa}: $S$ $\Sigma$ $s$ $F$ $\delta_{\mathit{dfa}}$ $\rightarrow$ $\mathit{|dfa|}$, \newline $( f \ b  \ t) \in \delta_{\mathit{dfa}}$ where $f, t \in S$ and $b \in \Sigma $
   \item \textbf{make-ndfa}: $S$ $\Sigma$ $s$ $F$ $\delta_{\mathit{ndfa}}$ $\rightarrow$ $\mathit{|ndfa|}$, \newline $( f \ b  \ t) \in \delta_{\mathit{ndfa}}$ where $f, t \in S$ and $b \in \{\Sigma \cup \{\epsilon\}\}$
   \item \textbf{make-pda}: $S$ $\Sigma$ $\Gamma$ $s$ $F$ $\delta_{\mathit{pda}}$ $\rightarrow$ $\mathit{|pda|}$, \newline $((f \ b  \ g) (t \ l)) \in \delta_{\mathit{pda}}$ where $f, t \in S$, $b \in \{\Sigma \cup \{\epsilon\}\}$, and \newline $g, l$ $\in$ $\{$\{non-empty list of $\Gamma$-symbols\} $\cup \ \{\epsilon\}\}$
   \item \textbf{make-tm}: $S$ $\Sigma$ $\delta_{\mathit{tm}}$ $s$ $F$ $\rightarrow$ $\mathit{|tm|}$, \newline $((f \ b) (t \ l)) \in \delta_{\mathit{tm}}$ where $f, t \in S$ and $b, l \in \{\Sigma \cup \{\epsilon, \sqcup\}\}$
\end{enumerate}
State machines are represented as functions that dispatch to the appropriate observer given some input.

\textsf{FSM} also employs a more general notion of a machine. A machine ($\mathit{m}$) is either:
 \begin{enumerate}
   \item A state machine
   \item Combined Turing machine (\textsf{ctm})\footnote{In terms of computational power there is no difference between a Turing Machine and a combined Turing machine. The latter is simply an abstraction that simplifies the design of Turing machines built from existing Turing Machines.}.
 \end{enumerate}
A \textsf{ctm} is not defined by providing a formal description. Instead, the user provides a \textsf{ctm} description ($ctmd$) and $\Sigma$. A \textsf{ctm} description uses other Turing machines as building blocks. In essence, a \textsf{ctm} represents an iterative algorithm with conditional branches and gotos that changes the state of the machine.  A $ctmd$ is described by the following grammar:\\ \\
\begin{centering}
\begin{tabular}{rcl}
  $<$$\mathit{ctmd}$$>$ & $\rightarrow$ & $\mathit{()}$ \\
   & $\rightarrow$ & ($<$\textsf{tm}$>$ $<$$\mathit{ctmd}$$>$) \\
   & $\rightarrow$ & ($<$$\mathit{ctmd}$$>$ $<$$\mathit{ctmd}$$>$) \\
   & $\rightarrow$ &  ($<$\textsf{label}$>$ $<$$\mathit{ctmd}$$>$) \\
   & $\rightarrow$ & (\textsf{BRANCH} \{$<$\textsf{symbol}$>$ $<$$\mathit{ctmd}$$>$\}$^+$) \\
   & $\rightarrow$ & (\textsf{GOTO} $<$\textsf{label}$>$) \\
   & $\rightarrow$ & ((\textsf{VAR} $<$symbol$>$) $<$$\mathit{ctmd}$$>$) \\
\end{tabular}
\end{centering}\\

\noindent Informally, the varieties of $\mathit{ctmd}$s are described as follows. The simplest $\mathit{ctmd}$s are either: empty, a \textsf{tm} followed by a $\mathit{ctmd}$, a $\mathit{ctmd}$ followed by a $\mathit{ctmd}$, or a label (used by a goto) followed by a $\mathit{ctmd}$. More complex $\mathit{ctmd}$s, can either start with a branch, a goto, or a variable. A branch defines options for the machine to execute based on the alphabet symbol being read. A goto unconditionally branches to a label that must be defined. Finally, to aid in the development of these machines, the library allows for branches to abstract over the symbol being read. A variable (not an alphabet symbol) captures the read symbol. This variable is used as a constant, whose value  can be written to the tape, in the \textsf{ctm} that follows it\footnote{In programming languages, this is akin to $\beta$-conversion.}.

The constructor signature for a \textsf{ctm} is: \textbf{combine-tms}: $ctmd$ $\Sigma$ $\rightarrow$ \textsf{ctm}. The observer \textbf{apply-ctm} takes as input a \textsf{ctm}, a tape configuration (akin to a word), and a head position and returns a Turing machine configuration (i.e., a state, a tape, and the position of the head on the tape). A \textsf{ctm} is represented as a function that returns a Turing machine configuration. In this manner, \textsf{ctm}$_1$ and \textsf{ctm}$_2$ can be composed as the resulting tape and head position of \textsf{ctm}$_1$ is used to build the initial configuration for \textsf{ctm}$_2$.

The state-machine transformers build new machines from existing machines, from grammars, or from a regular expression. These constructors implement algorithms from constructive proofs typically covered in an introductory formal languages course. They include:
\begin{description}
   \item [(regexp$\rightarrow$fsa $|\mathit{regexp}|$)] Transforms a regular expression into a finite-state machine.

   \item [(ndfa$\rightarrow$dfa $|\mathit{ndfa}|$)] Transforms a \textsf{ndfa} into a \textsf{dfa}.

   \item [(rename-states-sm $(\mathit{listof} \mathit{state})$ $|\mathit{sm}|$)] Renames the states of the given \textsf{sm} such that the intersection of the new names and the given list of states (i.e., symbols) is empty\footnote{In programming languages, this is akin to $\alpha$-conversion.}. This function is useful when combining two state machines requires that the intersection of the set of states of both machines be empty like, for example, when creating a machine using closure under union.

   \item [(union-sm $|\mathit{sm}|$ $|\mathit{sm}|$)] Builds an \textsf{sm} that accepts the union of the languages of the two given \textsf{sm}s of the same kind. If the inputs are \textsf{tm}s, they must be language recognizers.

   \item [(concat-sm $|\mathit{sm}|$ $|\mathit{sm}|$)] Builds an \textsf{sm} that accepts the concatenation of the languages of the two given \textsf{sm}s of the same kind. If the inputs are \textsf{tm}s, they must be a language recognizers.

   \item [(kleenestar-sm $|\mathit{sm}|$)] Builds an \textsf{sm} that accepts the Kleene star of the given \textsf{sm}'s language. If the input is a \textsf{tm}, it must be a language recognizer.

   \item [(complement-sm $|\mathit{sm}|$)] Builds an \textsf{sm} that accepts the complement of the language of the given \textsf{sm}. The given \textsf{sm} cannot be a \textsf{pda}. If the input is a \textsf{tm}, it must be a language recognizer.

   \item [(intersection-sm $|\mathit{sm}|$ $|\mathit{sm}|$)] Builds an \textsf{sm} that accepts the intersection of the languages of the two given \textsf{sm}s of the same kind. The given \textsf{sm}s cannot be \textsf{pda}s. If the inputs are \textsf{tm}s, they must be language recognizers.

   \item [(grammar$\rightarrow$sm $|\mathit{grammar}|$)] Builds an \textsf{sm} for the language of the given grammar.
\end{description}

The observers to return a given component of an \textsf{sm} take as input an \textsf{sm} and return the desired component. The interesting observers are:
\begin{description}
  \item [(apply-sm $|\mathit{sm}|$ $word$ $\{natnum\}$)] Runs the given \textsf{sm} assuming the given word is on the tape and the head of the \textsf{sm} is on position 0 of the tape. If the optional natural number is provided, the head starts at that position. The returned value is either \textquotesingle accept or \textquotesingle reject. The given \textsf{sm} is simulated by performing a breadth-first search of all the possible paths it can take by consuming the given word.

  \item [(show-transitions-sm $|\mathit{sm}|$ $word$ $\{natnum\}$)] Similar to \textsf{apply-sm}, but returns the path followed by the \textsf{sm}. For a nondeterministic \textsf{sm}, it returns an empty path if all possible paths end in a non-accepting state.
\end{description}

The testers for state machines offer the user the ability to experiment with the machines they create and provide immediate feedback. There are three testers provided:
\begin{description}
  \item [(same-result-sm? $|\mathit{sm}|$ $|\mathit{sm}|$ $word$)] Determines if both of the given \textsf{sm}s produce the same result for the given word.

  \item [(test-equiv-sm $|\mathit{sm}|$ $|\mathit{sm}|$ $\{natnum\}$)] Determines if both of the given \textsf{sm}s produce the same result on 100 randomly generated words. If the optional natural number is provided, then that number of random tests are performed. The function returns true or a list of words for which the given machines produce a different result. If a \textsf{tm} is provided as input, it must be a language recognizer.

  \item [(test-sm $|\mathit{sm}|$ $\{natnum\}$)] Generates 100 (or the given optional number of) random words and returns a list of the words with the results obtained from the given \textsf{sm}. If a \textsf{tm} is provided as input, it must be a language recognizer.
\end{description}

\subsection{Grammars}
A grammar is either:
 \begin{enumerate}
   \item Regular grammar (\textsf{rg})
   \item Context-free grammar (\textsf{cfg})
   \item Context-sensitive grammar (\textsf{csg})
 \end{enumerate}
Every grammar is composed of a set of terminal and nonterminal symbols ($V$), a set of terminal symbols ($\Sigma$), a set of of production rules ($R$), and a starting nonterminal symbol $S$. The left hand side of production rules for an \textsf{rg} and a \textsf{cfg} must only have a single nonterminal. The left hand side of production rules for a \textsf{csg} must contain at least one nonterminal. The right hand side of a production rule for a \textsf{rg} must have either a terminal symbol, a terminal symbol followed by a nonterminal symbol, or, if the left hand side is $S$, $\epsilon$ (the empty string). The right hand side of a production rule for a \textsf{cfg} and a \textsf{csg} may be $\epsilon$ or may have one or more members of $V$. The primitive constructors for grammars are:
\begin{enumerate}
  \item \textbf{make-rg}: $V$ $\Sigma$ $R_{\mathit{rg}}$ $S$ $\rightarrow$$|\mathit{rg}|$, \newline $(A \rightarrow aB) \vee (A \rightarrow a) \vee (S \rightarrow \epsilon) \in R_{\mathit{rg}}$ where $A, B \in V-\Sigma$ and $a \in \Sigma$

  \item \textbf{make-cfg}: $V$ $\Sigma$ $R_{\mathit{cfg}}$ $S$ $\rightarrow$$|\mathit{cfg}|$, \newline $(A \rightarrow a^+) \vee (A \rightarrow \epsilon) \in R_{\mathit{cfg}}$ where $A \in V-\Sigma$ and $a \in V$

  \item \textbf{make-csg}: $V$ $\Sigma$ $R_{\mathit{csg}}$ symbol $\rightarrow$ $|\mathit{csg}|$, \newline $BTC \rightarrow (D \vee \epsilon) \in R_{\mathit{csg}}$ where $B, C, D \in V^+$ and $T \in V-\Sigma$
\end{enumerate}

The transformers for grammars build new grammars from either existing grammars, existing state machines, or an existing regular expression. They include:
\begin{description}
  \item[(sm$\rightarrow$grammar $|sm|$)] Transforms a state machine into a grammar. If the input is a \textsf{tm}, it must be a language recognizer.

  \item[(grammar-rename-nts $(\mathit{listof} \mathit{symbol})$ $|grammar|$)] Renames the nonterminals of the given grammar to symbols not contained in the given list. This function is useful when the intersection of the nonterminals of two grammars must be empty.
\end{description}

The observers to return a given component of a grammar take as input a grammar and return the desired component. The interesting observer is:
\begin{description}
  \item [(deriv $|grammar|$ $word$)] If the given word is in the language of the given grammar, the derivation of the word is returned. Otherwise, a string stating that the word is not a member of the language of the grammar is returned.
\end{description}

The testers for grammars offer students the ability to experiment with the grammars they create and offer immediate feedback. There are three testers provided:
\begin{description}
  \item [(both-deriv? $|grammar|$ $|grammar|$ $word$)] Determines if both of the given grammars derive the given word.

  \item [(test-equiv-grammar $|grammar|$ $|grammar|$ $\{natnum\}$)] Determines if both of the given grammars produce the same result when attempting to derive 100 randomly generated words. If the optional natural number is provided, then that number of random tests are performed. The function returns true or a list of words for which the given grammars produce a different result.

  \item [(test-grammar $|grammar|$ $\{natnum\}$)] Generates 100 (or the given optional number of) random words and returns a list of the words with the results obtained from trying to use the given grammar to derive them.
\end{description}

\subsection{Regular Expressions}
Let $\Sigma$ be an alphabet of symbols. A regular expression ($re$) is a string  that is either
\begin{enumerate}
  \item $\epsilon$ (the empty string)
  \item $a \in \Sigma$
  \item $(re_1 \cup re_2)$, where $re_1$ and $re_2$ are $re$
  \item $(re_1re_2)$, where $re_1$ and $re_2$ are $re$
  \item $re^*$
\end{enumerate}
There is a constructor for each of the above. There is a single transformer, (\textbf{fsa$\rightarrow$regexp} $|fsa|$), that converts an \textsf{ndfa} to a regular expression. There is a single observer, (\textbf{printable-regexp} r), that takes as input a regular expression and that returns a string representing the given regular expression.

\section{The \textsf{FSM} Library in Practice}
This section presents examples of the \textsf{FSM} library in practice using problems that students have faced in an introductory formal languages course. The examples highlight how students and instructors make the use of the library relevant to both and how immediate feedback enhances the learning experience.

\subsection{Write a Program to Recognize a Regular Language}
\label{dfa}
Let $\Sigma = \{a, b\}$. Consider the problem of recognizing the regular language:
\begin{quote}
$L = \{$$w | w \in \Sigma^* \wedge$ \emph{w starts and ends with an a}$\}$.
\end{quote}
It is not uncommon for a student to submit the automaton in Figure \ref{SampleDFA1} which is clearly incorrect and earns the student a poor grade. The poor grade usually leads a student to experience frustration and apathy for the material. This frustration stems from realizing that the design has what they consider a small bug that experimentation and immediate feedback easily uncovers.

\newcommand{\DFA}{%

\begin{psmatrix}[rowsep=.5cm,colsep=.2cm,mnode=circle,fillstyle=solid,fillcolor=LightBlue]
             & & & & & & & & [name=N4] ds  \\
             & & & & [name=N1] q$_0$ \\
             & & & & & & & & [name=N2] q$_1$  & & & & & &  [name=N3] q$_2$\\
             & & & &
\end{psmatrix}

\psset{arrows=->,linecolor=black,arcangle=30,arrowsize=4pt
2,labelsep=2pt}}

\newcommand{\DFAc}{%

\begin{psmatrix}[rowsep=.5cm,colsep=.2cm,mnode=circle,fillstyle=solid,fillcolor=LightBlue]
             & & & & & & & & [name=N5] ds  \\
             & & & & [name=N6] q$_0$ \\
             & & & & & & & & [name=N7] q$_1$  & & & & & &  [name=N8] q$_2$\\
             & & & &
\end{psmatrix}

\psset{arrows=->,linecolor=black,arcangle=30,arrowsize=4pt
2,labelsep=2pt}}

\begin{figure}[t]
\begin{minipage}[b]{0.5\linewidth} 
\begin{tabular}{l}
  \DFA
  \rput(-4.47,2.1){\psframebox*[framearc=.1]{$>$}}
  \cnode(-.45,0.86){.28}{""}
  \ncarc {N1}{N4}
  \Aput{b}
  \ncarc{N2}{N3}
  \Aput{a}
  \ncarc{N3}{N2}
  \Aput{b}
  \psset{arrows=<-}
  \ncarc{N2}{N1}
  \Aput{a}
  \nccircle{->}{N4}{.35}
  \nbput{a,b}
  \nccircle{->}{N2}{.35}
  \nbput{b}
  \nccircle{->}{N3}{.30}
  \nbput{a}
\end{tabular}
\caption{Buggy Finite-State Automaton for $L$.}
\label{SampleDFA1}
\end{minipage}
\begin{minipage}[b]{0.5\linewidth} 
\begin{tabular}{l}
  \DFAc
  \rput(-4.48,2.1){\psframebox*[framearc=.1]{$>$}}
  \cnode(-2.37,0.86){.28}{""}
  \ncarc {N6}{N5}
  \Aput{b}
  \ncarc{N7}{N8}
  \Aput{b}
  \ncarc{N8}{N7}
  \Aput{a}
  \psset{arrows=<-}
  \ncarc{N7}{N6}
  \Aput{a}
  \nccircle{->}{N8}{.30}
  \nbput{b}
  \nccircle{->}{N5}{.35}
  \nbput{a,b}
  \nccircle{->}{N7}{.35}
  \nbput{a}
\end{tabular}
\caption{Correct Finite-State Automaton for $L$.}
\label{SampleDFA2}
\end{minipage}
\end{figure}

If the student has access to the \textsf{FSM} library, an implementation of their design looks as follows:
\begin{quote}
\begin{verbatim}
(define sol1-dfa (make-dfa  '(q0 q1 q2 ds)
                             '(a b)
                             'q0
                             '(q2)
                             '((q0 a q1)
                               (q0 b ds)
                               (q1 a q2)
                               (q1 b q1)
                               (q2 a q2)
                               (q2 b q1)
                               (ds a ds)
                               (ds b ds))))
\end{verbatim}
\end{quote}
The student can now test this machine as follows:
\begin{quote}
\begin{verbatim}
> (test-sm sol1-dfa)
'(((b b a a b b b b a b a b b b b a b) reject)
  ((b a b a a a b a b a a a a a b b b a b) reject)
  ((a a b a b a a b a a b b b b) reject)
  ((a b a a b b a b b b a) accept)
  ((a b b a) accept)
  ...
  ((a) reject)
  ...
  ((b a a a b b b b a a b b b a b) reject)
  ((b a b a a b b b) reject))
\end{verbatim}
\end{quote}
The tests reveal that the word \textsf{'(a)} is rejected, but this word is an element of $L$. The student can now refine the solution to the one displayed in Figure \ref{SampleDFA2}. The corresponding implementation is:
\begin{quote}
\begin{verbatim}
(define sol1-dfa (make-dfa  '(q0 q1 q2 ds)
                             '(a b)
                             'q0
                             '(q1)
                             '((q0 a q1)
                               (q0 b ds)
                               (q1 a q1)
                               (q1 b q2)
                               (q2 a q1)
                               (q2 b q2)
                               (ds a ds)
                               (ds b ds))))
\end{verbatim}
\end{quote}

This example may seem deceptively simple, but it does have a significant impact on a student's attitude. The immediate feedback provided by the \textsf{FSM} library assists students to find errors in their designs and to avoid the frustration of receiving a poor grade. Furthermore, the experience encourages both the student and the instructor resulting in an enhanced learning experience by eliminating the need to spend time on low-level bugs and allowing for more time to be dedicated to harder material.

\subsection{The Reverse of a Regular Language}
Proving that the reverse of a regular language, $L_{rev}$, is regular requires a constructive proof that students in an introductory formal languages course are commonly asked to write. Typically, students are confused about how to tackle this kind of problem. They understand that they must show how to build a finite-state automaton, $M_{rev}$, for $L_{rev}$, but feel frustrated if their algorithm is not correct. In a programming course, students can implement and experiment with their proposed solution. The same ought to be true for a course in formal languages and this is made possible by \textsf{FSM}.

\newcommand{\reverse}{%

\begin{psmatrix}[rowsep=.5cm,colsep=.2cm,mnode=circle,fillstyle=solid,fillcolor=LightBlue]
             [name=N10] q$_0$ & & & & &  [name=N11] q$_1$  & & & & & &  [name=N12] q$_2$ & & & & & &  [name=N13] ds \\
             & & & & & & & & & & & & & & & & & & & & & & &
\end{psmatrix}

\psset{arrows=->,linecolor=black,arcangle=30,arrowsize=4pt
2,labelsep=2pt}}
\begin{figure}[t]
\begin{center}
\begin{tabular}{l}
  \reverse
  \rput(-7.8,0.85){\psframebox*[framearc=.1]{$>$}}
  \cnode(-3.55,.85){.28}{""}
  \ncline{N11}{N12}
  \Aput{a}
  \ncline{N12}{N13}
  \Aput{a}
  \ncline{N10}{N11}
  \Aput{b}
  \psset{arrows=<-}
  \ncarc{N13}{N11}
  \Aput{b}
  \nccircle{->}{N10}{.30}
  \nbput{a}
  \nccircle{->}{N12}{.30}
  \nbput{b}
  \nccircle{->}{N13}{.30}
  \nbput{a,b}

\end{tabular}
\caption{A \textsf{dfa} for $L = a^*bab^*$.}
\label{Ldfsa}
\end{center}
\end{figure}

\newcommand{\reversec}{%

\begin{psmatrix}[rowsep=.7cm,colsep=.2cm,mnode=circle,fillstyle=solid,fillcolor=LightBlue]
             & & & & [name=N14] $S_n$ \\
             & & & & [name=N18] q$_0$ & & & & &  [name=N17] q$_1$  & & & & & &  [name=N15] q$_2$ & & & & [name=N16] ds \\
             & & & & & & & & & & & & & & & & & & & & & & &
\end{psmatrix}

\psset{arrows=->,linecolor=black,arcangle=30,arrowsize=4pt
2,labelsep=2pt}}

\newcommand{\reversef}{%

\begin{psmatrix}[rowsep=.7cm,colsep=.2cm,mnode=circle,fillstyle=solid,fillcolor=LightBlue]
             & & & & & & & & [name=N22] $S_n$ \\
            & & & &  & & & & [name=N19] q$_0$ & & & & &  [name=N20] q$_1$  & & & & & &  [name=N21] q$_2$
           \\
             & & & & & & & & & & & & & & & & & & & & & & &

\end{psmatrix}

\psset{arrows=->,linecolor=black,arcangle=30,arrowsize=4pt
2,labelsep=2pt}}

\begin{figure}[t]
\begin{minipage}[b]{0.5\linewidth} 
\begin{tabular}{l}
  \reversec
  \cnode(-6.4,1.055){.28}{""}
  \rput(-7.01,2.6){\psframebox*[framearc=.1]{$>$}}
  \ncline{N17}{N18}
  \nbput{b}
  \ncline{N14}{N15}
  \Aput{$\epsilon$}
  \ncline{N15}{N17}
  \nbput{a}
  \ncline{N16}{N15}
  \Aput{a}
  \ncarc{N16}{N17}
  \Aput{b}
  \nccircle{->}{N16}{.30}
  \nbput{a,b}
  \nccircle{->}{N15}{.30}
  \nbput{b}
  \nccircle{->}{N18}{.30}
  \nbput{a}
\end{tabular}
\caption{$M_{rev}$ from Theorem Version I.}
\label{reva}
\end{minipage}
\begin{minipage}[b]{0.5\linewidth} 
\begin{tabular}{l}
  \reversef
  \cnode(-4.875,1.055){.28}{""}
  \rput(-5.5,2.6){\psframebox*[framearc=.1]{$>$}}
  \ncline{N20}{N19}
  \nbput{b}
  \ncline{N22}{N21}
  \Aput{$\epsilon$}
  \ncline{N21}{N20}
  \nbput{a}
  \nccircle{->}{N21}{.30}
  \nbput{b}
  \nccircle{->}{N19}{.30}
  \nbput{a}
\end{tabular}
\caption{$M_{rev}$ from Theorem Version II.}
\label{revb}
\end{minipage}
\end{figure}

A common proposed solution is to build $M_{rev}$ from a deterministic finite-state automaton, $M$, for $L$. Intuitively, $M_{rev}$ has $M$'s set of states and a new starting state, $M$'s alphabet, $M$'s transition rules reversed,  $\epsilon$-transitions from the new starting state to each final state of $M$, and the starting state of $M$ as its only final state. More formally students define $M_{rev}$ in their proposed version I of the theorem as follows:
\begin{quote}
$M_{rev} = (\{Q_M \cup \{S_{rev}\}\}, \Sigma_M, \delta_{rev}, S_{rev}, \{S_M\})$, where
\begin{itemize}
  \item $Q_M = $ the states of $M$
  \item $S_{rev}$ is the unique symbol for the starting state of $M_{rev}$
  \item $\Sigma_M =$ the alphabet of $M$
  \item $\delta_{rev}$ contains two types of rules:
  \begin{enumerate}
     \item ($S_{rev}$, $\epsilon$, $q_i$) such that $q_i \in F_M =$ the final states of $M$
     \item ($q_i$, $\sigma$, $q_j$) such that ($q_j$, $\sigma$, $q_i$) $\in \delta_M$
  \end{enumerate}
  \item $S_M$ is the starting state of $M$
\end{itemize}
\end{quote}

The above algorithm is implemented and tested by students using \textsf{FSM}. Consider applying the proposed theorem to a deterministic finite-state automaton for $L = a^*bab^*$ depicted in Figure \ref{Ldfsa}. The resulting \textsf{ndfa} is depicted in Figure \ref{reva}. Students can observe that the state $ds$ is inaccessible and, therefore, ought not be part of the transformed machine. This observation suggests a refinement that eliminates the dead states of $M$ and any transitions involving a dead state. Using this refinement, the resulting \textsf{ndfa} for $M_{rev}$ is given in Figure \ref{revb}. An \textsf{FSM} implementation of this second version of the proposed theorem is displayed in Figure \ref{mrev}. Once testing makes students confident that their solution is correct, they can proceed to write the formal proof.

\begin{figure}[t]
\begin{verbatim}
  ; dfa --> ndfa
  (define (reverse-dfa m1)

    ;symbol (listof rules)--> boolean
    (define (deadstate? s rules)
      (let  ((fromrules (filter (lambda (r) (eq? s (smrule-fromstate r)))
                                rules)))
             (andmap (lambda (r) (eq? s (smrule-tostate r))) fromrules)))

    (let* ((newfinal (sm-getstart m1))
           (mstates (sm-getstates m1))
           (newStart(gen-symbol 'S mstates))
           (deadsts (remove-duplicates
                      (filter (lambda (a) (deadstate? a (sm-getdeltas m1)))
                              mstates)))
           (newQ (cons newStart (filter (lambda (q) (not (member q deadsts)))
                                mstates)))
           (addedrules (map (lambda (s) (list newStart EMP s))
                            (sm-getfinals m1)))
           (changedrules
             (map reverse
                  (filter (lambda (r) (and (not (member (smrule-tostate r)
                                                        deadsts))
                                           (not (member (smrule-fromstate r)
                          deadsts))))
                          (sm-getdeltas m1)))
          (newrules (append addedrules changedrules)))
      (make-ndfa newQ (sm-getsigma m1) newStart (list newfinal) newrules)))
\end{verbatim}
\caption{Proposed solution to build $M_{rev}$}
\label{mrev}
\end{figure}

Observe that in \textsf{FSM} new constructors are easily defined by students by simply writing a function. The code in Figure \ref{mrev} only uses primitive functions on lists, $\lambda$-expressions, and list-based higher-order functions (e.g., \textsf{map}, \textsf{filter}, and \textsf{andmap}). Thus, the coding of this constructor ought to be well within the reach of both advanced undergraduate and beginning graduate Computer Science students. Students uncomfortable or unfamiliar with $\lambda$-expressions and higher-order functions can, instead, omit their use by explicitly defining functions for $\lambda$-expressions and by explicitly defining recursive functions to perform the necessary list-processing done by the higher-order functions.

\subsection{Determining if a Context-Free Language is Empty}

\begin{figure}
\begin{verbatim}
  ; cfg --> boolean
  (define (Lcfg-isempty? g)
    ; cfg-rule (listof symbol) --> (listof cfg-rule)
    (define (only-accum-elems? rule accum)
      (and (not (member (cfg-rule-lhs rule) accum))
           (andmap (lambda (s) (member s accum)) (cfg-rule-rhs rule))))
    ; (listof cfg-rules) nonterminal (listof symbol) --> boolean
    (define (isempty? rls S accum)
      (cond [(member S accum) #f]
            [else
             (let ((newmembers (map cfg-rule-lhs
                                    (filter (lambda (r)
                                              (only-accum-elems? r accum))
                                            rls))))
               (cond [(empty? newmembers) #t]
                     [else (isempty? rls S (append newmembers accum))]))]))
    (let ((rls (cfg-get-the-rules g))
          (S (cfg-get-start g))
          (sigma (cfg-get-alphabet g)))
      (isempty? rls S (cons EMP sigma))))
\end{verbatim}
\caption{A Function to Determine if the Language of a CFG is Empty.}
\label{cfge}
\end{figure}

Students also face exercises to prove that a problem is decidable. If a problem is decidable, then there is an algorithm to determine if an instance of a given problem fulfills stipulated conditions. Students can propose an algorithm for decidability, but without the ability to test their solution they remain unsure about its validity. Testing their algorithm increases their confidence and brings their Computer Science education as programmers to bear in a formal languages course. We do not want students, however, to implement decidability algorithms using Turing machines. That would simply be too unwieldy. Instead, students ought to be able to use a library like \textsf{FSM} to implement and test their algorithms.

For example, consider the problem of deciding if the language, $L(G)$, of a context-free grammar, $G$, is empty. Students realize that $L(G)$ is not empty if there exists a derivation for any word formed by elements of the alphabet of $G$. Usually, they must be guided to realize that what they need is an algorithm to detect the existence of a derivation by creating any backward derivation to $G$'s starting symbol starting from the elements of the alphabet of $G$ and $\epsilon$. The algorithm accumulates $\Sigma$ and the left hand sides of rules (i.e., nonterminal symbols) whose right hand sides only contain symbols in the accumulator. If at any step, $G$'s starting symbol is in the accumulator then $L(G)$ is not empty. If at any step, there are no new symbols to add to the accumulator then $L(G)$ is empty. Otherwise, the new symbols are added to the accumulator and the process recursively proceeds.

Figure \ref{cfge} displays the implementation of this algorithm using the \textsf{FSM} library. Notice, that this code demonstrates that observers are easily added by simply writing a function. This observer only utilizes \textsf{FSM} provided functions and list-processing functions. Thus, once again, putting it well within the grasp of Computer Science graduate and advanced undergraduate students. Nonetheless, students tend to make mistakes at first and their algorithm requires refinement. The most common mistake is to not include $\epsilon$ in the initial value of the accumulator which is easily discovered once implemented.

\subsection{Computing with Turing Machines}
Although Turing machines are not the most attractive programming abstraction, it is important for students to understand their power and the reason the abstraction is less than attractive. The best way for students to begin to understand the power of Turing machines is to have them design Turing machines. Implementing formal descriptions of Turing machines in \textsf{FSM} is similar to developing the finite-state automatons in Section \ref{dfa}. Such descriptions are best for language recognizers and operations that do not involve assignment (i.e., altering of the tape). For instance, the following Turing machine moves the head to the right of the current position of the head.

\begin{verbatim}
(define RI (make-tm  '(s h)
                     '(I add1 sub1)
                      (list
                       (list (list 's 'I) (list 'h RIGHT))
                       (list (list 's 'add1) (list 'h RIGHT))
                       (list (list 's 'sub1) (list 'h RIGHT))
                       (list (list 's BLANK) (list 'h RIGHT)))
                      's
                      '(h)))
\end{verbatim}

Computations with Turing machines, however, may require assignment and, thus, care during their development. In such a setting, offering abstractions is critical  to keep students engaged. Many textbooks on formal languages develop a notation that is graphical and more transparent. In essence, the notation connects Turing machines, not states, using conditional branches and gotos. This allows for progressively more complex Turing machines to be designed from simpler Turing machines.

Consider the student having to add or subtract 1 from a non-zero unary number. The first step is to state the precondition and the postcondition for the Turing Machine. For example, the algorithm can be designed assuming the machine starts in the following configuration: ($\mathit{op}$ $\sqcup$ $number$ \underline{$\sqcup$}), where $\mathit{op}$ is either \textsf{add1} or \textsf{sub1}, $\sqcup$ denotes a blank space, and the head is on the first blank after the number. The machine stops in the following configuration: ($\mathit{op}$ $\sqcup$ $number$ \underline{$\sqcup$}), where $number$ is the result of the computation. Assume that in addition to \textsf{RI} above the following simpler machines are also defined:
\begin{quote}
\begin{description}
  \item [LI] Moves the head one space to the left.
  \item [I] Writes I to the tape.
  \item [BL] Writes $\sqcup$ to the tape.
  \item [RB] Moves the head to the first blank to the right of the head.
  \item [LB] Moves the head to the first blank to the left of the head.
\end{description}
\end{quote}

\begin{figure}[t]
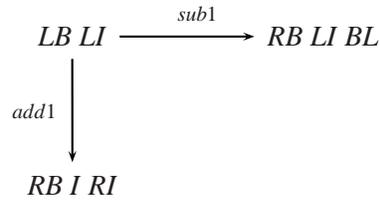

\[
\setlength{\arraycolsep}{1cm}
\def\tX{\tilde{\tilde{X}}}
\begin{array}{cc}
\Rnode{d}{LB\ LI} & \Rnode{e}{RB\ LI\ BL } \\[1.5cm]
\Rnode{f}{RB\ I\ RI}
\end{array}
\psset{nodesep=5pt,arrows=->}
\everypsbox{\scriptstyle}
\ncLine{d}{e}\Aput{sub1}
\ncLine{d}{f}\Bput{add1}
\]
\caption{Draft Turing Machine to add or subtract 1 from a non-zero unary number.}
\label{wrongtm}
\end{figure}
A first draft of the algorithm developed by a student is displayed in Figure \ref{wrongtm}. This algorithm moves the head to read the $\mathit{op}$. Then it branches depending on the $\mathit{op}$. After branching, it moves the head to the first blank to the right and proceeds to construct the resulting number. The \textsf{FSM} implementation is as follows:
\begin{verbatim}
(define addorsub (combine-tms (list LB LI (list BRANCH (list 'sub1 RB LI BL)
                                                       (list 'add1 RB I RI)))
                              '(I sub1 add1)))
\end{verbatim}
This machine can be tested using \textsf{apply-sm} to obtain the following results:
\begin{quote}
$>$ (apply-ctm $addorsub$ (list add1 \_ I I I I \_) 6)\\
(tmconfig 'h 2 '(add1 I I I I I \_))\\
$>$ (apply-ctm $addorsub$ (list sub1 \_ I I I I I I I \_) 9)\\
(tmconfig 'h 0 '(\_ \_ I I I \_))
\end{quote}
A student quickly realizes that the postcondition of the machine is not met and, therefore, their design has a bug. Instead of handing in a buggy design and being marked down by the instructor, the student can now proceed to redesign their algorithm. The bug, of course, is that after the branch the machine must move to the second blank to the right. The resulting graphical notation is displayed in Figure \ref{correcttm}. The \textsf{FSM} implementation is a s follows:

\begin{figure}[t]
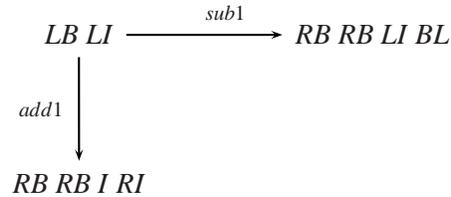

\[
\setlength{\arraycolsep}{1cm}
\def\tX{\tilde{\tilde{X}}}
\begin{array}{cc}
\Rnode{a}{LB\ LI} & \Rnode{b}{RB\ RB\ LI\ BL } \\[1.5cm]
\Rnode{c}{RB\ RB\ I\ RI} \\
\end{array}
\psset{nodesep=5pt,arrows=->}
\everypsbox{\scriptstyle}
\ncLine{a}{b}\Aput{sub1}
\ncLine{a}{c}\Bput{add1}
\]
\caption{Turing Machine to add or subtract 1 from a non-zero unary number.}
\label{correcttm}
\end{figure}
\begin{verbatim}
(define addorsub (combine-tms (list LB LI (list BRANCH (list 'sub1 RB RB LI BL)
                                                       (list 'add1 RB RB I RI)))
                              '(I sub1 add1)))
\end{verbatim}
The tests are repeated and the student gets the following results:
\begin{quote}
$>$ (apply-sm $addorsub$ (list add1 \_ I I I I \_) 6)\\
(tmconfig 'h 7 '(add1 \_ I I I I I \_))\\
$>$ (apply-sm $addorsub$ (list sub1 \_ I I I I I I I \_) 9)\\
(tmconfig 'h 8 '(sub1 \_ I I I I I I \_ \_))
\end{quote}
With successful tests, the student can now proceed to develop arguments for correctness. This development establishes that the feedback provided by testing Turing machines is an important part of the design process in a formal languages course.

\section{Related Work}
JFLAP \cite{Rodger} was designed to experiment with state machines and grammars as well as to experiment with constructive proofs. JFLAP allows the user to create and simulate several types of state machines, to create and parse strings in the language of a grammar, and to experiment with proof constructions such as converting a nondeterministic finite automaton to a deterministic finite automaton and then to a regular expression or regular grammar \cite{Rodger}. A study concluded that students felt more engaged and enjoyed a formal languages course more when using JFLAP \cite{Rodger}. JFLAP provides all the primitive and transformation constructors found in \textsf{FSM}. In contrast, however, the graphical nature of JFLAP does not offer the ability to easily add new observers and constructors to the software neither does JFLAP generate random testing.

The jFAST \cite{White} library assists beginners to design state machines. It uses a graphical interface just as JFLAP, but unlike JFLAP it provides no functionality for regular expressions and grammars. Like JFLAP and, in contrast to \textsf{FSM}, there is no support for students to add the constructive algorithms they develop and prove. The FSA Simulator allows the user to work and experiment with finite-state automata offering the ability to compare the languages of two finite-state automatons \cite{Grinder}. That is, it provides testing facilities for the equivalence of two finite-state automatons as \textsf{FSM}. The FSA comparison feature lets the software give students feedback about the accuracy of their work much as intended by \textsf{FSM} for all types of state machines. RegeXeX \cite{Brown} is an interactive system to write regular expressions. In contrast to \textsf{FSM}, it provides testing facilities for regular expressions. The feedback provided by RegeXeX includes strings that ought to be accepted and ought to be rejected much like \textsf{FSM} provides testing facilities for state machines.

\section{Conclusions and Future Work}
The \textsf{FSM} library provides users the necessary facilities to design and experiment with state machines, grammars, and regular expressions. It supports the view that the existence of a machine or grammar is proven by developing a constructive proof. This means that the proof presents an algorithm that can be implemented using \textsf{FSM}. Students and instructors no longer have to rely solely on paper-and-pencil traces to build confidence or discover bugs in a design. Instead, they can use \textsf{FSM} testing facilities to generate tests that provide immediate feedback. This leads students to actively reason and learn about formal languages as well as to reinforce their Computer Science education by implementing and developing unit tests for their constructive algorithms. The library has received positive feedback by students and has provided the examples presented in the article. It is our expectation that the proposed approach, namely teaching the theory of computation with tools to build computations, be widely adopted by Computer Science programs.

Future work includes expanding the library to include more constructors particularly those for state minimization. We will also extend the library to include a graphical interface. Unlike the interfaces described in the related work, we do not wish to have students create machines and grammars using a graphical interface. Instead, our goal is to have students continue to write code to create machines and grammars that are then rendered using graphics to animate execution and visualize their structure. Additionally, more support for regular expressions and extensions of Turing machines will be offered. The latter machines, although not computationally more powerful than a standard Turing Machine, are likely to make certain designs easier to implement by students. Finally, a goal is to convert the library into a pedagogy-friendly embedded DSL (e.g., using Racket's hygienic macros \cite{AutomataMacros}) where it is possible to machine check the proofs (e.g., using tools like DrACuLa \cite{Eastlund} and Coq \cite{Coq}).

\bibliographystyle{eptcs}
\bibliography{fsm}

\begin{thebibliography}{10}
\providecommand{\bibitemdeclare}[2]{}
\providecommand{\surnamestart}{}
\providecommand{\surnameend}{}
\providecommand{\urlprefix}{Available at }
\providecommand{\url}[1]{\texttt{#1}}
\providecommand{\href}[2]{\texttt{#2}}
\providecommand{\urlalt}[2]{\href{#1}{#2}}
\providecommand{\doi}[1]{doi:\urlalt{http://dx.doi.org/#1}{#1}}
\providecommand{\bibinfo}[2]{#2}

\bibitemdeclare{inproceedings}{Brown}
\bibitem{Brown}
\bibinfo{author}{Christopher~W. \surnamestart Brown\surnameend} \&
  \bibinfo{author}{Eric~A. \surnamestart Hardisty\surnameend}
  (\bibinfo{year}{2007}): \emph{\bibinfo{title}{RegeXeX: an interactive system
  providing regular expression exercises}}.
\newblock In: {\sl \bibinfo{booktitle}{SIGCSE '07: Proceedinds of the 38th
  SIGCSE technical symposium on Computer science education}},
  \bibinfo{publisher}{ACM Press}, \bibinfo{address}{New York, NY, USA}, pp.
  \bibinfo{pages}{445--449}, \doi{10.1145/1227310.1227462}.

\bibitemdeclare{inproceedings}{Eastlund}
\bibitem{Eastlund}
\bibinfo{author}{Carl \surnamestart Eastlund\surnameend} \&
  \bibinfo{author}{Matthias \surnamestart Felleisen\surnameend}
  (\bibinfo{year}{2009}): \emph{\bibinfo{title}{Automatic Verification for
  Interactive Graphical Programs}}.
\newblock In: {\sl \bibinfo{booktitle}{Proceedings of the Eighth International
  Workshop on the ACL2 Theorem Prover and Its Applications}},
  \bibinfo{series}{ACL2 '09}, \bibinfo{publisher}{ACM}, \bibinfo{address}{New
  York, NY, USA}, pp. \bibinfo{pages}{33--41}, \doi{10.1145/1637837.1637843}.

\bibitemdeclare{article}{Grinder}
\bibitem{Grinder}
\bibinfo{author}{Michael~T. \surnamestart Grinder\surnameend}
  (\bibinfo{year}{2003}): \emph{\bibinfo{title}{A Preliminary Empirical
  Evaluation of the Effectiveness of a Finite State Automaton Animator}}.
\newblock {\sl \bibinfo{journal}{SIGCSE Bull.}}
  \bibinfo{volume}{35}(\bibinfo{number}{1}), pp. \bibinfo{pages}{157--161},
  \doi{10.1145/792548.611958}.

\bibitemdeclare{article}{AutomataMacros}
\bibitem{AutomataMacros}
\bibinfo{author}{Shriram \surnamestart Krishnamurthi\surnameend}
  (\bibinfo{year}{2006}): \emph{\bibinfo{title}{EDUCATIONAL PEARL: Automata via
  Macros}}.
\newblock {\sl \bibinfo{journal}{J. Funct. Program.}}
  \bibinfo{volume}{16}(\bibinfo{number}{3}), pp. \bibinfo{pages}{253--267},
  \doi{10.1017/S0956796805005733}.

\bibitemdeclare{book}{Lewis}
\bibitem{Lewis}
\bibinfo{author}{Harry~R. \surnamestart Lewis\surnameend} \&
  \bibinfo{author}{Christos~H. \surnamestart Papadimitriou\surnameend}
  (\bibinfo{year}{1997}): \emph{\bibinfo{title}{Elements of the Theory of
  Computation}}, \bibinfo{edition}{2nd} edition.
\newblock \bibinfo{publisher}{Prentice Hall PTR}, \bibinfo{address}{Upper
  Saddle River, NJ, USA}.

\bibitemdeclare{book}{Linz}
\bibitem{Linz}
\bibinfo{author}{P.~\surnamestart Linz\surnameend} (\bibinfo{year}{2012}):
  \emph{\bibinfo{title}{An Introduction to Formal Languages and Automata}},
  \bibinfo{edition}{5th} edition.
\newblock \bibinfo{publisher}{Jones \& Bartlett Learning}.

\bibitemdeclare{book}{Martin}
\bibitem{Martin}
\bibinfo{author}{J.C. \surnamestart Martin\surnameend} (\bibinfo{year}{2003}):
  \emph{\bibinfo{title}{Introduction to Languages and the Theory of
  Computation}}.
\newblock \bibinfo{series}{McGraw-Hill Series in Computer Science},
  \bibinfo{publisher}{McGraw-Hill}.

\bibitemdeclare{manual}{Coq}
\bibitem{Coq}
\bibinfo{author}{\surnamestart \mbox{The Coq Development Team}\surnameend}
  (\bibinfo{year}{2004}): \emph{\bibinfo{title}{The Coq Proof Assitance
  Reference Manual}}.
\newblock \bibinfo{organization}{LogiCal Project}.
\newblock \urlprefix\url{http://coq.inria.fr}.
\newblock \bibinfo{note}{Version 8.0}.

\bibitemdeclare{misc}{Race}
\bibitem{Race}
\bibinfo{author}{Phil \surnamestart Race\surnameend} (\bibinfo{year}{2001}):
  \emph{\bibinfo{title}{Using Feedback to Help Students Learn}}.
\newblock
  \urlprefix\url{http://www.heacademy.ac.uk/resources/detail/resource_database%
/id432_using_feedback}.

\bibitemdeclare{article}{Rodger}
\bibitem{Rodger}
\bibinfo{author}{Susan~H. \surnamestart Rodger\surnameend},
  \bibinfo{author}{Eric \surnamestart Wiebe\surnameend},
  \bibinfo{author}{Kyung~Min \surnamestart Lee\surnameend},
  \bibinfo{author}{Chris \surnamestart Morgan\surnameend},
  \bibinfo{author}{Kareem \surnamestart Omar\surnameend} \&
  \bibinfo{author}{Jonathan \surnamestart Su\surnameend}
  (\bibinfo{year}{2009}): \emph{\bibinfo{title}{Increasing Engagement in
  Automata Theory with JFLAP}}.
\newblock {\sl \bibinfo{journal}{SIGCSE Bull.}}
  \bibinfo{volume}{41}(\bibinfo{number}{1}), pp. \bibinfo{pages}{403--407},
  \doi{10.1145/1539024.1509011}.

\bibitemdeclare{book}{Sipser}
\bibitem{Sipser}
\bibinfo{author}{Michael \surnamestart Sipser\surnameend}
  (\bibinfo{year}{2013}): \emph{\bibinfo{title}{Introduction to the Theory of
  Computation}}, \bibinfo{edition}{3rd} edition.
\newblock \bibinfo{publisher}{Cengage Learning}.

\bibitemdeclare{inproceedings}{White}
\bibitem{White}
\bibinfo{author}{Timothy~M. \surnamestart White\surnameend} \&
  \bibinfo{author}{Thomas~P. \surnamestart Way\surnameend}
  (\bibinfo{year}{2006}): \emph{\bibinfo{title}{jfast: A java finite automata
  simulator}}.
\newblock In: {\sl \bibinfo{booktitle}{In Thirty-seventh SIGCSE Technical
  Symposium on Computer Science Education}}, pp. \bibinfo{pages}{384--388},
  \doi{10.1145/1121341.1121460}.

\end{thebibliography}
\end{document}